\documentclass[runningheads]{llncs}
\usepackage{graphicx}
\usepackage{algorithm}
\usepackage{algpseudocode}
\usepackage{amsmath}
\usepackage{float}
\usepackage{multirow}
\usepackage{array}
\begin{document}
\title{A tiny public key scheme based on Niederreiter Cryptosystem}
%
%
\author{Arash Khalvan \and
Amirhossein Zali\and Mahmoud Ahmadian Attari}
\authorrunning{A. Khalvan et al.}
%
\institute{Department of Electrical Engineering, K. N. Toosi University of Technology, Tehran, Iran.
\email{\{a.khalvan,a.zali1\}@email.kntu.ac.ir}
\\  \email {mahmoud@eetd.kntu.ac.ir}}

\maketitle              
\begin{abstract}
Due to the weakness of public key cryptosystems encounter of quantum computers, the need to provide a solution was emerged. The McEliece cryptosystem and its security equivalent, the Niederreiter cryptosystem, which are based on Goppa codes, are one of the solutions, but they are not practical due to their long key length. Several prior attempts to decrease the length of the public key in code-based cryptosystems involved substituting the Goppa code family with other code families. However, these efforts ultimately proved to be insecure. 
\\In 2016, the National Institute of Standards and Technology (NIST) called for proposals from around the world to standardize post-quantum cryptography (PQC) schemes to solve this issue. After receiving of various proposals in this field, the Classic McEliece cryptosystem, as well as the Hamming Quasi-Cyclic (HQC) and Bit Flipping Key Encapsulation (BIKE), chosen as code-based encryption category cryptosystems that successfully progressed to the final stage.
\\This article proposes a method for developing a code-based public key cryptography scheme that is both simple and implementable. The proposed scheme has a much shorter public key length compared to the NIST finalist cryptosystems. The key length for the primary parameters of the McEliece cryptosystem (n=1024, k=524, t=50) ranges from 18 to 500 bits. The security of this system is at least as strong as the security of the Niederreiter cryptosystem. The proposed structure is based on the Niederreiter cryptosystem which exhibits a set of highly advantageous properties that make it a suitable candidate for implementation in all extant systems.

\end{abstract}
\section{Introduction}
Over the past few decades, cryptography has evolved from a collection of fundamental techniques and methods to become an integral component of communication systems. In addition to advancements in encryption techniques, a corresponding set of attack techniques has been developed to test the efficacy of contemporary security systems.
\\One of the most important works in the field of cryptography was the development of a method for secure information transmission without the need for a secret key agreement between the parties, the first example of which was introduced in 1976 by Diffie and Hellman in \cite{1}.
\\Most current cryptographic algorithms, such as RSA and ECC, rely on the challenge of integer factorization or discrete logarithm as their fundamental basis. RSA was initially presented to the public in 1978 by Rivest-Shamir and Adelman \cite{2}. 
\\ Following the publication of Shor's well-known paper in \cite{3}, which demonstrated the capability of quantum processors to break cryptosystems that rely on complex number theory problems like prime number factorization and discrete logarithms, one of the most significant threats to these systems emerged. This threat appears to undermine the fundamental security guarantees that cryptography aims to deliver.
\\ Due to the sudden growth of online businesses and their dependence on the Internet platform for many necessities of life such as banking, hospitals, Internet of Things, insurance, social networks, online stores, etc. Security threats are increasing. Additionally, the growth of investment and attention from large companies in quantum processing means that current cryptographic algorithms may be broken sooner than expected, leaving the security of any network uncertain.
\\In a recent study, Gidney \cite{5} demonstrated that the computation of the logical qubit required for the decomposition of an n-bit number can be derived from the $3n+0.002n$ lg $n$ relationship. Drawing upon reasonable assumptions regarding the relevant parameters, it is possible to estimate that the breaking of a 2048-bit RSA public key cryptography algorithm can be achieved within approximately eight hours, provided that a total of 20 million physical qubits are available. 
\\The possible threat of quantum attacks on the digital infrastructure protected by conventional cryptographic protocols has created an urgency to identify and implement countermeasures that can reduce the threat. Accordingly, there is a growing demand for more robust cryptographic schemes that integrate the advantages of both classical and quantum technologies. 
Post-quantum cryptography has emerged as a potential solution that can withstand the challenges posed by advances in quantum computing. Code-based encryption is a subset of PQC that was initially presented by R. McEliece in \cite{6}. However, H. Niederreiter later proposed a different encryption system that relied on solving a syndrome equations \cite{7}. This approach was essentially the dual of the McEliece scheme, and subsequent research demonstrated that both schemes offer an equivalent level of security \cite{8}.
\\The primary drawback of McEliece's cryptosystem is its large length of public key, which typically falls within the range of 50 to 100 KB. This feature has thus far hindered its practical implementation in real-world scenarios.
\\In response to the rising development of quantum processors, NIST initiated a competition in 2016 \cite{9}, inviting the cryptographic community to participate. The aim was to identify the most effective schemes, which would become the standard for cryptosystems. These schemes are referred to as post-quantum standard. Three code-based encryption schemes have reached the fourth and final round of the competition \cite{10}, which include the Classic McEliece, HQC, and BIKE encryption systems. The scheme introduced here is based on the Niederreiter  cryptosystem, which has a very short public key length.
This scheme is not only simple to use and implement, But it is also shown to have the minimum security of the original McEliece version and its Niederreiter equivalent. Its favorable features make it well-suited for use in a variety of systems.
\\ Many suggestions were made before the NIST competition to shorten the public key of the original McEliece cryptosystem by substituting the Goppa code family with different code families. However, each of these alternatives was eventually compromised and found to be an insecure choice.
For example, In 2007, Minder and Shokrollahi were able to break the system created by Sidelnikov \cite{11}, which was based on Reed-Muller codes, using a structural attack \cite{12}.
The authors of \cite{13} utilized the wild McEliece cryptosystem, which is based on wild Goppa codes, to create smaller public key sizes than the original McEliece cryptosystem. This was done in order to withstand all known attacks. However, in \cite{14}, it was demonstrated that this structure was not secure enough due to a polynomial-time structural attack.
\\In \cite{15}, Baldi and his colleagues developed a novel code-based cryptography system that relies on QC-LDPC codes. Similarly, in \cite{16}, they created another system based on QC-MDPC codes. However, both of these systems were found to be vulnerable to the OTD attack, which was proposed in \cite{17}.
In 2012, a new system that utilized convolutional codes was introduced \cite{18}. However, its vulnerability was discovered in \cite{19}, and it was subsequently broken using a structural attack. In 2014, there was a proposal for a public key cryptosystem that was based on polar codes \cite{20}. However, the security of this system was compromised in 2016 \cite{21}.
\\During 2016, NIST took a significant step towards the widespread use of PQC by requesting proposals for standardizing PQC schemes. The focus was on three main areas: public key encryption (PKE), digital signatures (DS) and key encapsulation mechanisms (KEM) \cite{9}. These areas are recognized as public key cryptographic tools and are evaluated based on three criteria: security, cost and implementation.
\\ Many algorithms from around the world claiming post-quantum security have been submitted to NIST. But only a few of them reached to the final stage and we are waiting for the final standardization of these encryption systems. NIST plans to choose a group of cryptographic algorithms that belong to various families of cryptographic systems, instead of selecting a single winner. This approach aims to minimize the risk of cryptanalysis of the chosen algorithm and provide alternative options in case of any issues \cite{22}. NIST qualified a total of 69 papers in this call and reviewed the various candidates that advanced from the first round to the second round \cite{23}.
\\After passing four stages of the competition, Finally 3 code base encryption systems reached the final stage.
The final stage of the competition saw three encryption systems make it through: Classic McEliece, which utilizes the Goppa code in the Niederreiter cryptosystem and offers a high level of IND-CCA2 security \cite{24}; Bit flipping Key Encapsulation, which employs Quaci cyclic MDPC codes \cite{25}; and Hamming Quasi-Cyclic, which is based on quasi-cyclic codes \cite{26}. The final winners will be chosen by 2024. So far, four schemes have emerged victorious in previous competitions, while the rest have been eliminated \cite{27,28}. Table~\ref{tab1} provides a summary of the latest results.
\begin{table}[t]
\centering 

\caption{Summary of the results of the fourth round of NIST standardization \cite{28}}\label{tab1}
\renewcommand{\arraystretch}{1.5} 
\begin{tabular}{|>{\centering\arraybackslash}p{7cm}|>{\centering\arraybackslash}p{2.5cm}|c|}\hline
\multicolumn{1}{|c|}{\bfseries Problem}        & \multicolumn{1}{c|}{ \bfseries Status}                     & \bfseries Algorithm                                                                                                                              \\ \hline
\multicolumn{3}{|c|}{Public-key   encryption and key exchange group}                                                                                                                                                            \\ \hline
\multicolumn{1}{|c|}{Code-based}     & \multicolumn{1}{c|}{Candidate   for round 4}    & Classic   McEliece                                                                                                                     \\ \hline
\multicolumn{1}{|c|}{Lattice-based}  & \multicolumn{1}{c|}{Winner,   becomes standard} & CRYSTALS-KYBER                                                                                                                         \\ \hline
\multicolumn{1}{|c|}{Lattice-based}  & \multicolumn{1}{c|}{Withdrawn}                  & NTRU                                                                                                                                   \\ \hline
\multicolumn{1}{|c|}{Lattice-based}  & \multicolumn{1}{c|}{Withdrawn}                  & SABER                                                                                                                                  \\ \hline
\multicolumn{1}{|c|}{Code-based}     & \multicolumn{1}{c|}{Candidate   for round 4}    & BIKE                                                                                                                                   \\ \hline
\multicolumn{1}{|c|}{Lattice-based}  & \multicolumn{1}{c|}{Withdrawn}                  & FrodoKEM                                                                                                                               \\ \hline
\multicolumn{1}{|c|}{Code-based}     & \multicolumn{1}{c|}{Candidate   for round 4}    & HQC                                                                                                                                    \\ \hline
\multicolumn{1}{|c|}{Lattice-based}  & \multicolumn{1}{c|}{Withdrawn}                  & NTRU   Prime                                                                                                                           \\ \hline
\multicolumn{1}{|c|}{Isogeny-based}  & \multicolumn{1}{c|}{Candidate   for round 4}    & SIKE                                                                                                                                   \\ \hline
\multicolumn{3}{|c|}{Digital signatures group}                                                                                                                                                                                  \\ \hline
\multicolumn{1}{|c|}{Lattice-based}  & \multicolumn{1}{c|}{Winner,   becomes standard} & CRYSTALS  DILITHIUM\\ \hline
\multicolumn{1}{|c|}{Lattice-based}  & \multicolumn{1}{c|}{Winner,   becomes standard} & FALCON                                                                                                                                 \\ \hline
\multicolumn{1}{|c|}{Multivariate}   & \multicolumn{1}{c|}{Withdrawn}                  & Rainbow                                                                                                                                \\ \hline
\multicolumn{1}{|c|}{Multivariate}   & \multicolumn{1}{c|}{Withdrawn}                  & GeMSS                                                                                                                                  \\ \hline
\multicolumn{1}{|c|}{Zero-knowledge} & \multicolumn{1}{c|}{Withdrawn}                  & Picnic                                                                                                                                 \\ \hline
\multicolumn{1}{|c|}{Hash-based}     & \multicolumn{1}{c|}{Winner,   becomes standard} & SPHINCS+                                                                                                                               \\ \hline
\end{tabular}
\end{table}
\\ Although significant efforts have been made, the risk of a quantum attack still persists. This is due to the fact that there are both old and new devices that must be transferred to the PQC algorithm set, which will take several years to transition from the current system to the secure PQC set.
\\In \cite{29}, the main and appropriate strategies to protect systems against quantum attacks and methods to combine current cryptography with PQC to minimize potential damage are discussed, and provide a suitable perspective of PQC transition so that organizations can achieve a smooth and timely PQC transition.

\section{Proposed Cryptosystem}
The system proposed in this article, henceforth denoted as Kal1, is a public key encryption system that is very similar to the Niederreiter cryptosystem~\cite{7}. This system performs encryption and decryption exactly according to the Niederreiter cryptosystem. However, the primary distinction between Kal1 and the Niederreiter is the construction of their public keys. Specifically, Kal1's public key is significantly shorter, with a reduction in length exceeding 99.8\%.
 \subsection{Proposed Cryptosystem Algorithm}
 Similar to all public key systems, this cryptosystem comprises a set of three distinct algorithms: key generation, encryption and decryption.
\begin{algorithm}
\caption{Key generation}\label{alg:1}
\begin{algorithmic}
\Require $System\;parameters:\;$n, m and\;t.$ $
\Ensure $The\;cyclic\;public\;key\;{p_k}=( {H^T}_{cyclic},\;t)\;and\;the\; private \;key\;{s_k}=(S_{ns},H,P_{per}).$
\State$1:\;Choose\;a\;randomly\;sequence\;({\alpha _0},{\alpha _1}, \cdots ,{\alpha _{n - 1}})\;of\;n\; distinict\;elements\;in\;GF({2^m}).$
\State$2:Choose \,a\;random\;polynomial \;g(x)\;such\;g(\alpha ) \ne 0\;for\;all\;\alpha  \in ({\alpha _0},{\alpha _1}, \cdots ,{\alpha _{n - 1}}).$
\State$3:\;Compute\;the\;t\times n\;parity\;check\;matrix: $
\State$H = \left[ {\begin{array}{*{20}{c}}
{1/g({\alpha _0})}&{1/g({\alpha _1})}& \cdots &{1/g({\alpha _{n - 1}})}\\
{{\alpha _0}/g({\alpha _0})}&{{\alpha _1}/g({\alpha _1})}& \cdots &{{\alpha _{n - 1}}/g({\alpha _{n - 1}})}\\
 \vdots & \vdots & \ddots & \vdots \\
{\alpha _0^{t - 1}/g({\alpha _0})}&{\alpha _1^{t - 1}/g({\alpha _1})}& \cdots &{\alpha _{n - 1}^{t - 1}/g({\alpha _{n - 1}})}
\end{array}} \right]$.
\State $4:\; Choose\;a\;random\;vector\;and\;shift\;it\;in\;k\;iteration\;to\;create\; a\;cyclic\;matrix\; P_{cyclic}. $
\State$\hspace{0.5cm}\;Padding\;identity\;matrix\;to\;P_{cyclic}\;in\;order\;to\;calculate\;H_{cyclic}:$
\State$\hspace{0.5cm}\;H_{cyclic} = {[{P_{cyclic}}|{I_{(n - k) \times (n - k)}}]_{(n - k) \times n}}.$
\State $5:\;Generate\;randomly\;an\;inevitable\;matrix\;S_{ns} \in {M_{k,k}}({F_2})\;in\;F_2. $
\State $6:\;Generate\;randomly\;a\;permutation\;matrix\;P_{per} \in {M_{n,n}}({F_2})\;in\;F_2. $
\State $7:\;Calculate\;{{H'}^T}\;from\;{{H'}^T} = {P_{per}^T}{H^T}{S_{ns}^T}.$
\State $8:\;Calculate\ H_{{\rm{secondary}}}^T\;from\;H_{{\rm{secondary}}}^T = {H^T}_{cyclic} + {{H'}^T}.$
\State $9:\;\textbf{return: } \;{s_k} ,\;{p_k}.$
\end{algorithmic}
\end{algorithm}
\begin{algorithm}
\caption{Encryption}\label{alg:2}
\begin{algorithmic}
\Require $The \;cyclic\;public \;key \;{p_k}=( {H^T}_{cyclic},\;t)\;and \;a\;message\;to\;encrypt\;m \in F_2^k.$
\Ensure $c \in F_2^n\;Cipher\;text\;associated\;with\;m.$
\State $1:\; Mapping\;m\;to\;a\;vector\;of\;size\;n-k\;and\;weight\;t:$
\State$\hspace{0.5cm}\; e_i = \varphi (m).$
\State $2:\;Padding\;a\;zero\;vector\;with\;size\;1 \times k\;before\;e_i :$
\State$\hspace{0.5cm}\;e = {[Zero{s_{1 \times k}}|{e_i}]_{1 \times n}}.
$
\State $3:\;encrypt \;the\;message\;c = e \times {H^T}_{cyclic}.$
\State $4:\;\textbf{return: } \;{c} .$
\end{algorithmic}
\end{algorithm}
\begin{algorithm}
\caption{Decryption}\label{alg:3}
\begin{algorithmic}
\Require $The \;private \;key \;{s_k} = (S_{ns},H,P_{per})\;and \;cipher\;text\;to\;decrypt \;c \in F_2^n.$
\Ensure $m \in F_2^k\;the\;clear\;text\;associated\;with\;c.$
\State $1:\;Calculate\;{c_1} = c \times {({S_{ns}^T})^{ - 1}}.$
\State$2:\;Syndrome\;decoding\;{c_1}:\;{c_2} = Decode({c_1}).$
\State $4:\;Calculate\;{c_3} = {c_2} \times {({P_{per}^T})^{ - 1}}=e= {[zero{s_{1 \times k}}|{e_i}]_{1 \times n}}.$
\State $5:\;Calculate\;m = {\varphi ^{ - 1}}(e_i).$
\State $6:\;\textbf{return: } \;{m} .$
\end{algorithmic}
\end{algorithm}

\begin{figure}[h]
 \centering
  \includegraphics[width=0.3\linewidth]{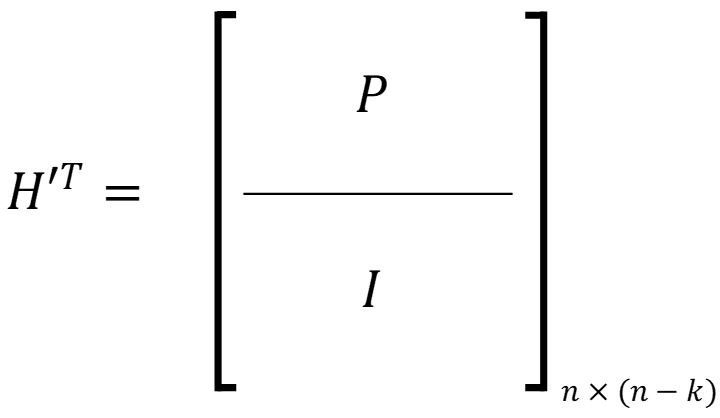}
  \caption{${{H'}^T}$ matrix structure.}
  \label{fig:my_labe2}
\end{figure}

\begin{figure}[h]
 \centering
  \includegraphics[width=0.3\linewidth]{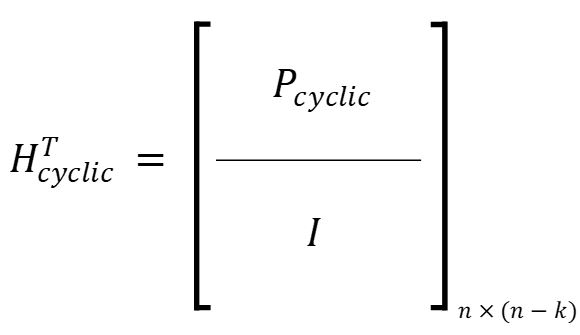}
  \caption{$H_{cyclic}^T$ matrix structure.}
  \label{fig:my_labe3}
\end{figure}

\begin{figure}[H]
 \centering
  \includegraphics[width=0.3\linewidth]{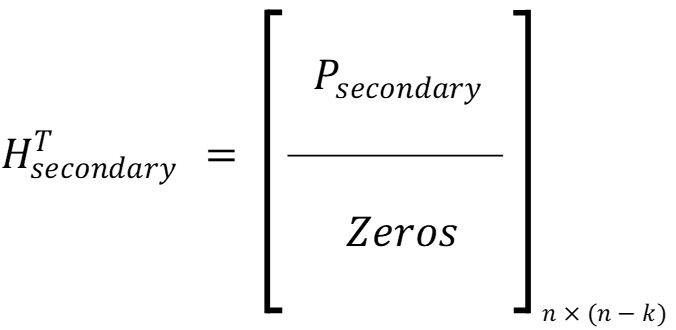}
  \caption{$H_{secondary}^T$ matrix structure.}
  \label{fig:my_labe4}
\end{figure}

\newpage The following Fig. ~\ref{fig:my_label} summarizes the proposed system algorithms when transmitting a message m from Alice to Bob.
 \begin{figure}
  \includegraphics[width=\linewidth]{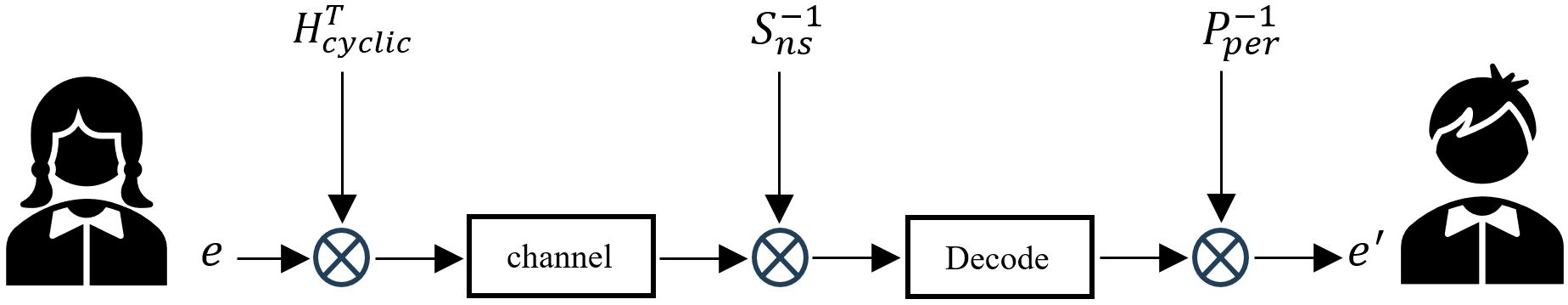}
  \caption{Block diagram of the proposed system}
  \label{fig:my_label}
\end{figure}
\\
In public key encryption systems, the working method is as follows: first, a public key is shared, with which the encryption operation is performed. The private keys, which possess the capability to decrypt the encrypted message, are exclusively accessible to authorized receiver.
\\
In the proposed system, Bob has shared the public key $H_{cyclic}^T$ which is in the equation ~\ref{eq:emc1} with all network users. Alice is now able to send Bob an encrypted message using his public key.
\begin{equation}
 H_{cyclic}^T = {{H'}^T} + H_{secondary}^T 
\label{eq:emc1}
\end{equation}
The matrix denoted as ${{H'}^T}$ is precisely identical to the Niederreiter systematic encryption matrix, as evidenced by equation ~\ref{eq:emc2} and Fig.~\ref{fig:my_labe2}:
\begin{equation}
{{H'}^T} = {P_{per}^T}{H^T}{S_{ns}^T}
\label{eq:emc2}
\end{equation}
\\ To generate the ${{H'}^T}$ it is necessary for Bob to select a binary Goppa code that is generated by a matrix ${G_{k \times n}}$ and has the capability to correct $t$ errors. Additionally, he randomly selects two matrices, $S_{ns}$ and $P_{per}$, where $S_{ns}$ is an invertible matrix and $P_{per}$ is a permutation matrix. Subsequently, Bob calculated ${{H'}^T}$.
\\ As depicted in Fig.~\ref{fig:my_labe3}, the matrix $H_{cyclic}^T$ is composed of two distinct components, namely $I$ and ${P}_{cyclic}$. 
The ${P}_{cyclic}$ matrix is formed by first choosing a random vector with a length of $n-k$, which is then placed in the top row of the ${P}_{cyclic}$ matrix. The remaining rows are created by circularly shifting the first row in $k$ iterations.
\\The matrix $I$ is an identity matrix.
\\ $H_{secondary}^T$ matrix is computed by applying the equation $H_{secondary}^T = {{H'}^T} + H_{cyclic}^T$. As depicted in Fig.~\ref{fig:my_labe4}, The $H_{secondary}^T$ matrix is composed of two distinct components, namely ${P_{secondary}}$ and $Zeros$. The $Zeros$ component is a matrix consisting entirely of zero elements, while ${P_{secondary}}$ is computed by applying the equation ${P_{secondary}} = {P_{cyclic}} + P$.

\begin{figure}
 \centering
  \includegraphics[width=0.7\linewidth]{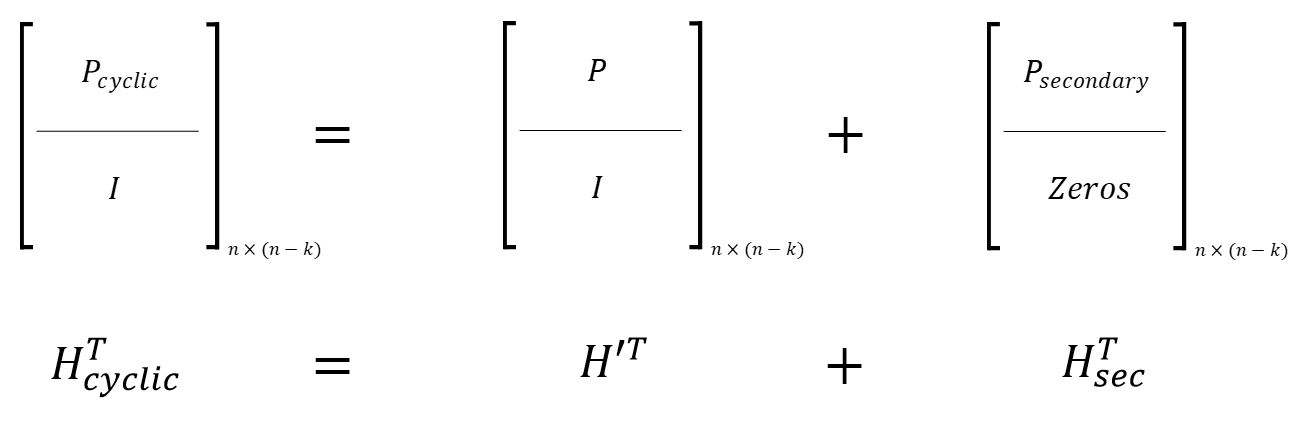}
  \caption{Comprehensive structure of $H_{cyclic}^T$ matrix.}
  \label{fig:my_labe5}
\end{figure}

\begin{figure}
 \centering
  \includegraphics[width=0.45\linewidth]{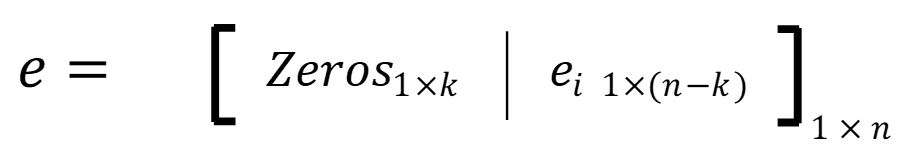}
  \caption{message structure (error vector)}
  \label{fig:my_labe6}
\end{figure}

\begin{figure}
 \centering
  \includegraphics[width=0.85\linewidth]{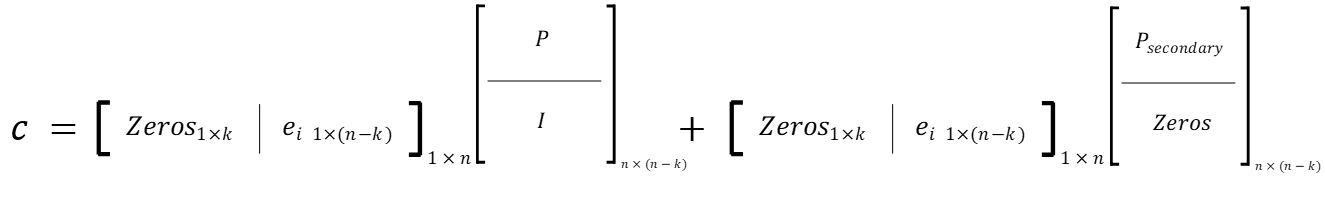}
  \caption{Encryption overview}
  \label{fig:my_labe7}
\end{figure}

A comprehensive illustration of the $H_{cyclic}^T$ matrix is presented in Fig.~\ref{fig:my_labe5}.
\\ Bob then publishes $H_{cyclic}^T$ as a public key and keep the rest of the secret information. Upon obtaining Bob's public key, Alice is able to transmit an encrypted message to Bob. However, prior to sending the message, Alice is required to perform a mapping of each message $m_i$  to the corresponding error vectors $e_i$, which possess a length of $n-k$ and a weight of $t$, in accordance with the structure depicted in Fig.~\ref{fig:my_labe6}. 
\\ Where $Zero{s_{1 \times k}}$ is a vector consisting entirely of zero elements.
\\ Now, Alice can send the encrypted message $c = e \times H_{cyclic}^T$ to Bob. According to the structure of $e$ and $H$, the encrypted message of $c$ will be in the form of equation~\ref{eq:emc3} and Fig.~\ref{fig:my_labe7}:
\begin{equation}
c = e \times H_{cyclic}^T = e \times {{H'}^T} + e \times H_{secondary}^T
\label{eq:emc3}
\end{equation}

According to Fig.~\ref{fig:my_labe6}, it is evident that the outcome of the second component of the $e \times H_{secondary}^T$ equation  is consistently equivalent to zero: 
 \begin{equation}
 c = e \times H_{cyclic}^T = e{P_{per}^T}{H^T}{S_{ns}^T}
\label{eq:emc4}
\end{equation}
 \\ Equation~\ref{eq:emc4} represents the encryption algorithm used in the Niederreiter cryptosystem, and it can be decrypted using the same method. Specifically, the decryption process is as follows:
\\ 1. Calculate ${c_1} = c \times {({S_{ns}^T})^{ - 1}}$:
 \begin{equation}
 {c_1} = e{P_{per}^T}{H^T}{S_{ns}^T}{({S_{ns}^T})^{ - 1}} = e{P_{per}^T}{H^T}
\label{eq:emc5}
\end{equation}
\\ 2. Syndrome decoding ${c_1}$:
 \begin{equation}
 {c_2} = Decode(e{P_{per}^T}{H^T}) = e{P_{per}^T}
\label{eq:emc6}
\end{equation}
\\ 3. Calculate ${c_3} = {c_2} \times {({P_{per}^T})^{ - 1}}$:
 \begin{equation}
 {c_3} = e{P_{per}^T}{({P_{per}^T})^{ - 1}} = e'
\label{eq:emc7}
\end{equation}
\\ 4. Recover message from remapping ${c_3}$:
 \begin{equation}
 m = {\varphi ^{ - 1}}({e_i})
\label{eq:emc8}
\end{equation}
Within the proposed cryptosystem, Bob  share only the first row of matrix $P_{cyclic}$ with the Alice. In order to construct $H_{cyclic}^T$, Alice can simply shift the received vector and append an identical matrix to it. In this case, the length of the public key will be equal to $n-k=500$ bits.
\\ Given the parameters of the original McEliece cryptosystem, specifically $(n,k,t)=(1024,524,50)$, it can be observed that the length of the public key is reduced by over 99.8\% when compared to the original versions of both the original McEliece and the Niederreiter cryptosystems.
Table~\ref{tab2} shows that the key size in the Kal1 cryptosystem has been shortened enough to be used in various systems.
\subsubsection {$Kal1-S1$}
In a special case, referred to as Kal1-S1 henceforth, Bob has the option to designate the first row of his matrix $P_{cyclic}$ as highly sparse, with a weight of 10, for instance.
In this case, it is only necessary to share the positions of each arrays that is equal to 1 as a public key. assuming considering the original McEliece design with parameters (1024,524,50), the first row of $P_{cyclic}$ has equal to $n-k=500$ ($n-k < 2^9$) bits, which requires 9 bits to address each 1 of the first row. In this case, the public key will be shorter ($90\times10=90$ bits).
\subsubsection {$Kal1-S2$}
\begin{figure}
 \centering
  \includegraphics[width=0.5\linewidth]{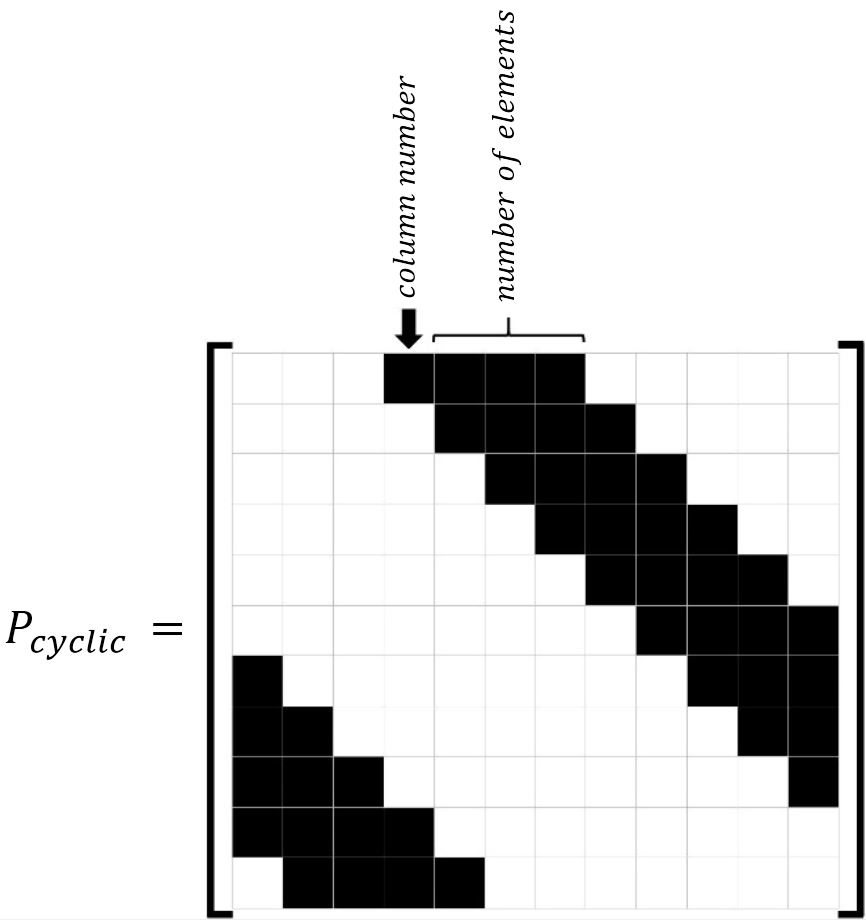}
  \caption{Kal1-S2 public key.}
  \label{fig:my_labe8}
\end{figure}

In a another special case, referred to as Kal1-S2 henceforth, If  Bob  considers 1's arrays of the first rows of $P_{cyclic}$  matrix consecutively, the length of the key will be shorter. In this situation, it is enough to specify the address of the first 1 array and the count of 1s that follow it, As depicted in Fig.~\ref{fig:my_labe8}. By doing this, the resulting key length will be significantly reduced.
 \begin{equation}
 \begin{array}{l}
{P_{compress}} = \{ \begin{array}{*{20}{c}}
{\begin{array}{*{20}{c}}
{\begin{array}{*{20}{c}}
{\begin{array}{*{20}{c}}
{column}&{number}
\end{array}|}&{number}
\end{array}}&{of}
\end{array}}&{elements}
\end{array}\} \\
\;\;\;\;\;\;\;\;\;\;\;\;\;\;\; {P_{compress}} = \{ 4|3\}  = \{ 0100|011\} 
\end{array}
\label{eq:emc9}
\end{equation}
\\According to the formula presented in Equation ~\ref{eq:emc9}, the Kal-S2 public key denoted by ${P_{compress}}$ reaches its shortest length in Kal1. This length will be the shortest feasible state for sharing a public key, represented by a vector, with a weight greater than 2.

As part of NIST's ongoing endeavors to use post-quantum cryptosystems, three code-based designs have successfully advanced to the final round of the competition in the latest round of NIST competitions \cite{24}.
Table \ref{tab2} makes a comparison between the proposed schemes and the NIST final round schemes. In this article, significantly shorter key than the introduced and final NIST schemes is presented \cite{28}.
\begin{table}
\centering 
\caption{A comparison between the proposed schemes and the NIST final round schemes.}\label{tab2}
\renewcommand{\arraystretch}{1.5} 
\begin{tabular}{|>{\centering\arraybackslash}p{5cm}|>{\centering\arraybackslash}p{2.5cm}|c|}
\hline
\bfseries Cryptosystem          & \bfseries ID       & \bfseries Public key lenght (bit) \\ \hline
Classic McEliece      & -        & 536576                  \\ \hline
Niederreiter          & -        & 262000                  \\ \hline
\multirow{2}{*}{BIKE} & BIKE\_L1 & 1541                    \\ \cline{2-3} 
                      & BIKE\_L3 & 3083                    \\ \hline
\multirow{3}{*}{HQC}  & HQC128   & 2289                    \\ \cline{2-3} 
                      & HQC192   & 4522                    \\ \cline{2-3} 
                      & HQC256   & 7245                    \\ \hline
proposed scheme       & Kal1     & 500                     \\ \hline
proposed scheme       & Kal1-S1  & 90                      \\ \hline
proposed scheme       & Kal1-S2  & 18                      \\ \hline
\end{tabular}
\end{table}
\newpage
\subsection{Implementation analysis}
As it was said, the encryption and decryption of the Kal1 scheme exactly match the Niederreiter design and the only difference between these two schemes is in the key generation step.
\\The Kal1 scheme employs a key generation relationship expressed as $H_{cyclic}^T = {{H'}^T} + H_{secondary}^T$. The sole variation in the key generation step is the binary addition of the $H_{secondary}^T$ matrix with the ${{H'}^T}$ matrix of the Niederreiter scheme. This operation can be executed by any type of hardware, and similar implemented schemes are documented in \cite{30,31,32,33} and \cite{34}.

\subsection{The security of Kal1 scheme}
 Consider the structure of the  $H_{cyclic}^T = {{H'}^T} + H_{secondary}^T$. The attacker only has $H_{cyclic}^T$ and needs the private keys of the Niederreiter scheme for decryption.
It can be posited that the security level of the Kal1 scheme is at least equivalent to the Niederreiter cryptosystem. However, the complexity of an attack on the Kal1 scheme would be greater for an attacker than that on the Niederreiter cryptosystem. In the Niederreiter cryptosystem, the process of obtaining the answer involves a sequential testing of private keys. However, in the Kal1 scheme, the process is more intricate. Prior to the stage of conjecturing private keys, the attacker must first be capable of separating $H_{cyclic}^T$ into its constituent parts, namely $H_{secondary}^T$ and ${{H'}^T}$.
\\The Niederreiter cryptosystem and its McEliece security equivalent, both of which rely on the Goppa code, have been subjected to security analysis for different types of attacks in \cite{35}. In this scheme, it is necessary to consider the challenge of separating $H_{cyclic}^T$ into its constituent parts.
\\ To separate $H_{cyclic}^T$ into its constituent parts, it is necessary to solve a system of equations involving $(n-k)\times n$ and $2\times(n-k)\times n$ unknowns, provided that the $H_{secondary}^T$  and ${{H'}^T}$ matrices are linearly independent in the binary space ${F_2}$. While it is not esay, but it can be solved. In the situation that the Bob can select $H_{cyclic}^T$  in a manner that results in one of the two matrices $H_{secondary}^T$ and ${{H'}^T}$ becoming linearly dependent, the resulting equation will not possess a unique solution \cite{36}.
\\ Typically, the Information-Set Decoding (ISD) attack \cite{37} involves the identification of an information set $I$, which is a subset of the set of columns of matrix $G$, denoted by $I \subseteq \{ 1,2,...,n\}$. Upon identifying an information set, an invertible sub-matrix ${\left( {{G_I}} \right)_{k \times k}}$ is constructed. Subsequently, $G_I^{ - 1}G$ is obtained as a systematic generating matrix for $C$, subject to the condition of permuting the columns. The vector $m$ is revealed by the positions of the information set $I$ in $C = mG_I^{ - 1}G$. In the Kal1 scheme, according to the equation~\ref{eq:emc10}, there is no guarantee for the full rank of the encryption matrix $H_{cyclic}^T$, and even the Bob can intentionally design  $H_{cyclic}^T$ in such a way that the full rank matrix with dimensions $k\times k$ cannot be found.
 \begin{equation}
rank(A + B) \le rank(A) + rank(B)
\label{eq:emc10}
\end{equation}
For ISD attack, this scheme can be more resistant than Niederreiter and Mceliece cryptosystems. Due to repel this type of attacks, $H_{secondary}^T$ and ${{H'}^T}$ matrix must be extract from $H_{cyclic}^T$. ISD attacks for the Niederreiter cryptosystem have been investigated in \cite{38,39,40} and\cite{41}.
\section{Conclusion}
Code-based cipher schemes have not been appropriate for structures with limited processing resources because they require a significant amount of memory, ranging from 64 KB to over 1 MB. The article presents a novel approach that has the smallest public key length ever introduced. Apart from being simplicity, this technique can be utilized in various systems and provides a minimum level of security equivalent to Niederreiter's cryptosystem.

%
%
%
\bibliography{ref}






\end{document}